\documentclass[final,3p,times,onecolumn]{elsarticle}
\onecolumn
\usepackage{hyperref}
\usepackage{pifont}
\usepackage{color}
\usepackage{xcolor}
\usepackage{tabularx, booktabs}
\usepackage{amsmath}
\usepackage{algorithmic}
\usepackage{float}
\usepackage{multirow}
\usepackage{algorithm}
\usepackage{caption}
\usepackage{fancyhdr}
\usepackage{natbib}
\usepackage{graphicx}
\newcommand{\cmark}{\ding{51}}%
\newcommand{\xmark}{\ding{55}}%

\begin{document}

\begin{frontmatter}

\title{SDC-GON: Singular Decomposition and Consistency-Regularized Green's Operator Networks for Solving Partial Differential Equations}

\author{Yingchao Huang\corref{mycorrespondingauthor}}
\ead{huangyi@saskpolytech.ca}
\address{Faculty of Digital Innovation, Arts \& Sciences, Saskatchewan Polytechnic, Regina SK S4S 5X1, Canada}
\author{Xin Wang}
\address{Faculty of Digital Innovation, Arts \& Sciences, Saskatchewan Polytechnic, Regina SK S4S 5X1, Canada}
\author{Shanshan Yao}
\address{Civil and Environmental Engineering, University of Alberta, Edmonton AB T6G 2H5, Canada}
\author{Fanhua Zeng}
\address{Energy Systems Engineering, Faculty of Engineering and Applied Science, University of Regina, SK S4S 0A2, Canada}
\author{Wei Peng}
\address{Industrial Systems Engineering, Faculty of Engineering and Applied Science, University of Regina, SK S4S 0A2, Canada}

\cortext[mycorrespondingauthor]{Corresponding author}

\begin{abstract}
Green's function based operator approximation offers an efficient route for solving linear partial differential equations under varying boundary conditions and source terms. Once the Green's function is learned, solutions for new configurations are obtained through integration rather than by solving the differential equation again. Existing Green's function learning methods face two structural challenges. The first is the singular behavior of the Green's function near the source point, which places a difficult approximation burden on neural networks. The second is the absence of explicit consistency between the learned Green's function and its gradient, although both quantities enter the integral solution representation directly. This work proposes SDC-GON, a Singular Decomposition and  Consistency-Regularized Green's Operator Network that addresses both challenges within a unified framework. The Green's function is decomposed into an analytically known singular component and a smooth correction learned by the network, so that the neural approximation targets only the regular part of the response  kernel. A self-consistency loss enforces  agreement between the gradient and the  autodifferentiation gradient of the smooth correction. The method is evaluated on two dimensional Poisson, three dimensional heat conduction, heterogeneous reaction diffusion, and Stokes benchmarks, consistently outperforming the compared baselines across all cases. On the heterogeneous pipe benchmark, SDC-GON achieves a testing error of $3.70\times10^{-4}$ with a smaller network architecture, compared with $9.60\times10^{-4}$ for the same-width baseline and $4.63\times10^{-4}$ for a larger configuration, demonstrating that structural improvements are more effective than increasing model size.
\end{abstract}

\end{frontmatter}

\section{Introduction}

Partial differential equations (PDEs) provide a central mathematical framework for modeling systems whose states vary across space, time, and material configuration \cite{evans2010partial,quarteroni2009numerical}. They arise in fluid dynamics, heat transfer, structural mechanics, and reaction diffusion processes, where conservation laws, constitutive relations, transport mechanisms, and boundary interactions must be represented in forms suitable for analysis and simulation \cite{batchelor2000introduction,murray2002mathematical}. This role makes PDEs foundational in computational science because many physical and engineering simulations require the prediction of solution fields governed jointly by differential operators, source terms, boundary conditions, and material coefficients.

Classical numerical methods provide the main computational foundation for solving these problems. Finite difference methods approximate differential operators through local difference stencils, finite element methods use variational formulations on discretized domains, and finite volume methods enforce conservation over control volumes \cite{leveque2007finite,zienkiewicz2024finite,eymard2000finite}. These methods are robust and widely adopted because they provide systematic discretizations for structured and unstructured domains. They are also supported by mature theories of stability, convergence, conservation, and approximation \cite{brenner2008mathematical}. Their main limitation appears when repeated forward evaluations are required under varying boundary conditions or source terms \cite{PeirO2005}. Such repeated evaluations occur frequently in parameter optimization, uncertainty quantification, and inverse problems \cite{tarantola2005inverse,stuart2010inverse,smith2013uncertainty}. In these settings, each new configuration generally requires a full independent numerical solve, which can make large scale exploration computationally expensive.

This repeated solve bottleneck motivates analytical representations that separate the response of the governing operator from the specific choice of source term and boundary condition. Green's functions provide such a representation for linear PDEs by describing the impulse response of the governing operator \cite{stakgold2011green}. Once the Green's function is known, the solution for a new source term and boundary condition can be obtained through an integral representation rather than by solving the differential equation again \cite{gu2025explainable}. In a discretized setting, this means that new solutions can be evaluated through quadrature using the same response kernel, instead of assembling and solving a new system for every input configuration. This reusability is the key appeal of Green's function based operator approximation. The fundamental challenge is that analytical Green's functions are generally tractable only for simple geometries and constant coefficient operators, while variable coefficient operators on complex domains rarely admit closed form expressions \cite{stakgold2011green}.

Deep learning has introduced several computational approaches for solving PDEs and accelerating repeated prediction tasks \cite{karniadakis2021physics}. These approaches can be grouped into solution approximation methods, operator learning methods including Green's function learning methods, and physics constrained operator learning methods. Solution approximation methods represent the solution of a given problem instance and are exemplified by physics informed neural networks \cite{raissi2019physics}. Operator learning methods instead learn mappings between input functions and solution fields, as in DeepONet and Fourier neural operators \cite{lu2021learning,li2021fourier}. Green's function learning methods form a physically structured branch of operator learning because they learn response kernels associated with linear systems \cite{boulle2022data}. Physics constrained operator learning methods further incorporate governing equation residuals into operator learning objectives, as in physics informed DeepONets \cite{wang2021learning}. This operator level viewpoint is especially relevant for the present work because the goal is not to solve one fixed PDE instance, but to evaluate many source terms and boundary conditions after training.

Within this context, recent progress in Green's function based operator approximation is represented by the Green's Operator Network (GON) \cite{gu2025explainable}. GON learns Green's functions on three dimensional bounded domains and evaluates solutions through Green's function integral representations. This design enables boundary invariant and source term invariant generalization for linear PDEs. Despite this progress, two limitations remain. First, the singular behavior of the Green's function near the source point is still learned directly by the neural network, which places a difficult non smooth approximation burden on the model. Second, the Green's function and its gradient are learned without explicit consistency enforcement, although both quantities enter the integral solution formula. These limitations are closely connected. When the learned function retains the algebraic singularity, automatic differentiation near the source diagonal can produce unstable gradient estimates for the learned approximation, which makes direct consistency enforcement on the raw Green's function unreliable. The present work therefore proposes Singular Decomposition and Consistency-Regularized Green's Operator Network (SDC-GON) to address the singularity and consistency limitations within a unified framework.

The first contribution of SDC-GON is a singular decomposition strategy in which the Green's function is separated into an analytically prescribed singular component and a smooth correction learned by the neural network. This follows the classical numerical analysis principle that singular kernels should be treated analytically or with specialized numerical techniques before the regular remainder is approximated \cite{sauter2011boundary,duffy1982quadrature}. By assigning the leading singular behavior to a known analytical term, the network is asked to learn a smoother correction rather than the full singular Green's function. Building on this decomposition, the second contribution is a self consistency loss imposed between the automatic differentiation gradient of the smooth correction and the directly predicted gradient. This closes the consistency gap while avoiding unstable gradient matching on the raw singular function \cite{baydin2018automatic}. These two components convert the original task from learning a singular Green’s function and an unconstrained gradient into learning a smooth correction and a consistent smooth gradient.

The remainder of the paper is organized as follows. Section~\ref{related_work} reviews related work, Section~\ref{methodology} presents the methodology, Section~\ref{results} reports experimental results, and Section~\ref{conclusion} concludes with a discussion of limitations and future directions.

\section{Related Work}
\label{related_work}

Machine learning methods for PDEs have developed from single instance solution approximation toward operator level prediction across families of related problems. This shift is important for the present study because SDC-GON is designed to evaluate new boundary conditions and source terms after training, rather than solve one fixed configuration at a time. The related literature is therefore reviewed through five connected themes: neural solution approximation, general neural operator learning, Green's function based operator learning, singularity treatment in learned Green's functions, and physics constrained operator learning.

Solution approximation methods use neural networks to represent the solution field of a particular PDE problem. Physics informed neural networks (PINNs) are representative of this category because they train neural networks by penalizing mismatch with the governing equation and the boundary or initial conditions \cite{raissi2019physics}. This formulation is attractive because physical constraints can be incorporated directly into the learning objective \cite{raissi2019physics}. It reduces dependence on dense labeled solution data in forward and inverse problems . Related mesh free methods include the deep Galerkin method \cite{sirignano2018dgm} and the deep Ritz \cite{e2018deep} method. The deep Galerkin method trains neural networks through domain sampling for high dimensional differential equations, while the deep Ritz method minimizes variational energy functionals for suitable variational problems. These methods provide flexible solution representations, but the trained network is usually tied to a particular problem instance or a particular training configuration \cite{WANG2020108963}. Substantial changes in source terms, boundary conditions, or material coefficients can therefore require retraining or additional adaptation. Optimization difficulties have also been documented for complex or stiff physics informed losses \cite{krishnapriyan2021characterizing}. These limitations make solution approximation methods less efficient for the repeated query setting considered in this paper.

Operator learning methods address this limitation by learning mappings between input functions and output solution functions \cite{NEURIPS2023_df543023}. DeepONet is a representative neural operator architecture because it uses a branch network to encode input functions and a trunk network to encode evaluation locations, enabling the trained model to evaluate different inputs from a learned operator family \cite{lu2021learning}. Fourier neural operators (FNOs) are another influential operator learning framework and are included as a baseline in this study \cite{li2021fourier}. FNO parameterizes the kernel integration operation in Fourier space by applying learned linear transformations to Fourier modes of the input representation, which makes it efficient for parametric PDEs on regular discretizations \cite{li2021fourier}. More recent neural operator studies have extended the operator learning viewpoint to broader function space settings and irregular discretizations, including message passing neural PDE solvers for mesh based representations \cite{kovachki2023neural,brandstetter2022message}. These general operator learning approaches support repeated prediction after training, but their learned representations are not explicitly the Green's function of the governing system. As a result, the response kernel is less directly interpretable in the classical analytical sense, and additional handling is often required when the data are defined on unstructured tetrahedral and boundary meshes, as in the GON benchmarks \cite{gu2025explainable}.

Green's function learning methods form a more physically structured branch of operator learning because they approximate the response kernel associated with a linear system. Boulle et al. \cite{boulle2022data} used rational neural networks to discover Green's functions from data and demonstrated that learned kernels can provide interpretable solution representations in low dimensional settings. Teng et al. \cite{teng2022learning} learned Green's functions for linear reaction diffusion equations by replacing the impulse source with a smooth surrogate in a residual based formulation. Dai et al. \cite{dai2024numerical} proposed a deep learning framework with Green's functions for boundary value problems, which further shows the relevance of Green's function representations for neural solution construction. Aldirany et al. \cite{aldirany2024operator} developed a Green's function based deep learning approach for wave equation operator approximation, extending this direction beyond elliptic model problems. DeepGreen also provides an important precedent by using learned Green's function representations in a framework for nonlinear boundary value problems through a learned linearization viewpoint \cite{gin2021deepgreen}. These studies show that Green's function learning can connect neural prediction with classical integral solution theory.

Recent work has further expanded Green's function learning toward more reusable and operator oriented representations. Negi et al. \cite{negi2024learning} proposed a boundary integral network to learn a domain independent Green's function for elliptic PDEs, which emphasizes reuse across domains and boundary conditions. Melchers et al. \cite{melchers2026neural} introduced neural Green's operators for parametric PDEs, where the operator construction is derived from the Green's formulation and is used to improve generalization across parametric inputs. GON advanced this direction by learning Green's functions in three dimensional bounded domains and using numerical integration to handle varying boundary conditions and source terms \cite{gu2025explainable}. The binary structured neural network used in GON was originally proposed to improve the representation of rapidly changing solutions and localized features, which makes it relevant for Green's function learning where diagonal behavior is difficult to approximate \cite{LIU2024113341}. These advances motivate SDC-GON, but they also reveal that the singular structure of the Green's function and the consistency between the learned scalar kernel and its gradient remain unresolved.

The singular behavior of Green's functions near the source point creates a specific challenge for Green's function learning. Classical numerical analysis does not usually approximate such singular kernels as ordinary smooth functions. Instead, known singular behavior is treated analytically or through specialized quadrature transformations before the remaining regular part is approximated \cite{duffy1982quadrature,monegato1994numerical}. Existing learning based Green's function methods address this issue in different but mostly indirect ways. Rational neural networks increase the flexibility of the approximating function class, smooth impulse surrogates make residual based training feasible, and binary structured neural networks improve the representation of localized features \cite{boulle2022data,teng2022learning,LIU2024113341}. Sun et al. \cite{sun2026learning} recently proposed singularity encoded Green's function learning, which embeds singularity related information into a neural learning formulation for Green's functions and iterative solvers. This is closely related in motivation to SDC-GON. The distinction is that SDC-GON uses an explicit analytical singular subtraction before neural approximation, so the network is trained on the smooth correction rather than on a singular or singularity encoded kernel. The proposed decomposition therefore serves two purposes: it stabilizes training by removing the singular component from the learned target, and it embeds the local structure of the Green's function analytically into the operator learning framework.

The treatment of singularity is closely connected to the way physical constraints are imposed during learning. Physics constrained operator learning methods combine the repeated prediction capability of operator learning with residual constraints from the governing equations \cite{SARTHAK2026114261}. PI-DeepONet is representative of this category and is used as a baseline in the present experiments \cite{wang2021learning}. It extends DeepONet by adding physics based residual constraints to the training objective, which can improve physical consistency and reduce reliance on labeled solution data \cite{wang2021learning}. This additional structure is valuable, but residual evaluation and automatic differentiation increase training cost during optimization \cite{karniadakis2021physics}. In contrast, GON \cite{gu2025explainable} and SDC-GON encode the linear operator structure through the Green's function integral representation rather than through a generic residual penalty.

SDC-GON further strengthens this structure by combining analytical singular decomposition with consistency regularization on the learned smooth correction. This consistency term differs from standard PDE residual losses used in physics constrained operator learning. Rather than enforcing the governing equation at collocation points, it enforces agreement between the gradient obtained by automatic differentiation of the learned smooth correction and the gradient predicted by the dedicated gradient head. This distinction is important because the boundary integral depends directly on the Green's function gradient, so inconsistency between the scalar Green's function representation and its gradient can produce incoherent boundary contributions. This positioning explains the role of each comparative method in the experiments: PINN represents solution approximation, DeepONet and FNO represent general operator learning, PI-DeepONet represents physics constrained operator learning, and GON represents the closest Green's function based operator learning baseline.

\section{Methodology}
\label{methodology}

This section presents the proposed SDC-GON framework for learning Green's function based solution operators for linear PDEs. The method first establishes the linear boundary value problem and the corresponding Green's function representation. It then introduces a singular decomposition that separates the analytically known local singularity from a smooth correction learned by the network. Based on this decomposition, SDC-GON uses a shared BsNN backbone with dual output heads to predict the smooth correction and its source coordinate gradient, and the two outputs are coupled through a self consistency loss. The resulting learned quantities are inserted into the same volume and boundary integral structure as GON, with Gaussian quadrature applied to the neural components and the closed form singular kernel evaluated explicitly within the same quadrature framework.

\subsection{Problem formulation}

Let $\Omega \subset \mathbb{R}^{d}$ be a bounded domain with boundary $\partial \Omega$. In this work, the following linear boundary value problem with Dirichlet boundary condition is considered:
\begin{equation}
\begin{cases}
\mathcal{L}(u)(\mathbf{x}) = f(\mathbf{x}), & \mathbf{x}\in \Omega,\\
u(\mathbf{x}) = g(\mathbf{x}), & \mathbf{x}\in \partial\Omega,
\end{cases}
\label{eq:pde_problem}
\end{equation}
where $u(\mathbf{x})$ is the solution field, $f(\mathbf{x})$ is the source term, $g(\mathbf{x})$ is the prescribed boundary value, and $\mathcal{L}$ is a linear differential operator. The main scalar cases considered in this work correspond to diffusion or diffusion--reaction type operators, where $a(\mathbf{x})$ denotes the diffusion coefficient and any reaction contribution is absorbed into the linear operator $\mathcal{L}$. The Green's function $G(\mathbf{x},\boldsymbol{\xi})$ is defined as the response at the evaluation point $\mathbf{x}$ induced by an impulse source located at $\boldsymbol{\xi}$. For homogeneous Dirichlet boundary conditions, it satisfies
\begin{equation}
\begin{cases}
\mathcal{L}(G)(\mathbf{x},\boldsymbol{\xi}) = \delta(\mathbf{x}-\boldsymbol{\xi}), & \mathbf{x}\in \Omega,\\
G(\mathbf{x},\boldsymbol{\xi}) = 0, & \mathbf{x}\in \partial\Omega,
\end{cases}
\label{eq:green_definition}
\end{equation}
where $\delta$ is the Dirac delta distribution. Once $G(\mathbf{x},\boldsymbol{\xi})$ is known, the solution of Eq.~\eqref{eq:pde_problem} can be evaluated by the Green's function integral representation
\begin{equation}
u(\mathbf{x})
=
\int_{\Omega}
f(\boldsymbol{\xi})G(\mathbf{x},\boldsymbol{\xi})\,d\boldsymbol{\xi}
-
\int_{\partial\Omega}
g(\boldsymbol{\xi})a(\boldsymbol{\xi})
\left(
\nabla_{\boldsymbol{\xi}}G(\mathbf{x},\boldsymbol{\xi})\cdot \mathbf{n}_{\boldsymbol{\xi}}
\right)
\,dS(\boldsymbol{\xi}),
\qquad
\mathbf{x}\in \Omega,
\label{eq:green_representation}
\end{equation}
where $\mathbf{n}_{\boldsymbol{\xi}}$ is the outward unit normal on $\partial\Omega$ at the boundary point $\boldsymbol{\xi}$. This representation is the basis of both GON and SDC-GON: once the Green's function and its boundary-normal derivative are available, new source terms and boundary values can be evaluated through the volume and boundary integrals rather than through a new PDE solve.

\subsection{Singular decomposition}

The direct approximation of $G(\mathbf{x},\boldsymbol{\xi})$ is difficult because the Green's function is singular when the evaluation point approaches the source point \cite{RICKAYZEN2003129}. For the Laplacian operator in three dimensions, the leading singular behavior is proportional to the reciprocal distance between $\mathbf{x}$ and $\boldsymbol{\xi}$ \cite{stakgold2011green}. Its gradient with respect to the source coordinate has stronger local variation, with magnitude proportional to the reciprocal squared distance \cite{ZHAN2025521}. Near the diagonal set of $\Omega\times\Omega$, the function learned by the network therefore changes rapidly and becomes unbounded in the limiting analytical model.

Neural network approximation theory shows that network complexity is closely related to the regularity of the target function, with more efficient approximation available for smoother function classes \cite{yarotsky2017error}. This makes direct learning of algebraically singular kernels less favorable than learning smooth corrections after the singular part has been removed. BsNN \cite{LIU2024113341} was introduced in GON \cite{gu2025explainable} to improve the representation of local features and the diagonal singular behavior of Green's functions. The need for this specialized architecture provides additional motivation for treating the singular component analytically rather than requiring the network to learn the full raw Green's function. SDC-GON addresses this difficulty by removing the known leading singular structure before applying neural approximation, so that the network learns only the smooth correction.

SDC-GON decomposes the Green's function into an analytically prescribed singular term and a smooth learnable correction:
\begin{equation}
G(\mathbf{x},\boldsymbol{\xi})
=
G_{\mathrm{sing}}(\mathbf{x},\boldsymbol{\xi})
+
G_{\mathrm{sm}}(\mathbf{x},\boldsymbol{\xi}),
\label{eq:green_decomposition}
\end{equation}
where $G_{\mathrm{sing}}$ denotes the leading singular component and $G_{\mathrm{sm}}$ denotes the smooth correction. The corresponding gradient decomposition is
\begin{equation}
\nabla_{\boldsymbol{\xi}}G(\mathbf{x},\boldsymbol{\xi})
=
\nabla_{\boldsymbol{\xi}}G_{\mathrm{sing}}(\mathbf{x},\boldsymbol{\xi})
+
\nabla_{\boldsymbol{\xi}}G_{\mathrm{sm}}(\mathbf{x},\boldsymbol{\xi}).
\label{eq:gradient_decomposition}
\end{equation}
For three dimensional diffusion dominated operators, the singular component is defined as
\begin{equation}
G_{\mathrm{sing}}(\mathbf{x},\boldsymbol{\xi})
=
\frac{1}
{4\pi\|\mathbf{x}-\boldsymbol{\xi}\|},
\label{eq:singular_green}
\end{equation}
which corresponds to the free-space singular kernel of the three dimensional Laplacian. This term is used to remove the dominant local singular behavior that appears as the evaluation point $\mathbf{x}$ approaches the source point $\boldsymbol{\xi}$.Coefficient effects are retained in the integral representation through the material coefficient in the boundary term and are further represented by the smooth correction $G_{\mathrm{sm}}$. This choice gives a simple universal singular subtraction that is shared across the diffusion dominated cases, while allowing geometry, boundary effects, and coefficient variation to be learned by the regular component. 


It should be noted that the singular component adopted in this work,
$G_{\mathrm{sing}}=1/(4\pi\|\mathbf{x}-\boldsymbol{\xi}\|)$, corresponds to the free-space Green's function of the constant-coefficient three-dimensional Laplacian. For the variable-coefficient diffusion--reaction operator in Eq.~(2), the leading local singularity is instead proportional to
$1/(4\pi a(\boldsymbol{\xi})\|\mathbf{x}-\boldsymbol{\xi}\|)$. Therefore, when $a(\boldsymbol{\xi})\neq 1$, the constant-coefficient singular subtraction does not fully remove the leading singularity of the variable-coefficient Green's function. The resulting correction term can retain a residual singular component of order $O(1/\|\mathbf{x}-\boldsymbol{\xi}\|)$, with coefficient proportional to $1/a(\boldsymbol{\xi})-1$. Thus, in heterogeneous cases, the decomposition used here should be interpreted as an approximate singular decomposition rather than a fully coefficient-adapted decomposition.

The gradient of the singular term is available in closed form and is written as
\begin{equation}
\nabla_{\boldsymbol{\xi}}G_{\mathrm{sing}}(\mathbf{x},\boldsymbol{\xi})
=
\frac{\mathbf{x}-\boldsymbol{\xi}}
{4\pi\|\mathbf{x}-\boldsymbol{\xi}\|^{3}}.
\label{eq:singular_gradient}
\end{equation}
The remaining correction $G_{\mathrm{sm}}$ accounts for the effects of domain geometry, boundary interaction, heterogeneous coefficients, and deviations from the free-space Laplacian singular model. Since the leading singular behavior has already been extracted analytically, $G_{\mathrm{sm}}$ is a smoother target for neural approximation. This changes the learning problem from approximating a singular kernel to approximating a regular correction, which is better aligned with the approximation properties of neural networks.

For the two-dimensional Poisson benchmark, the singular component was defined using the free-space Green's function of the two-dimensional Laplace operator. Specifically, for $\mathbf{x},\boldsymbol{\xi}\in\mathbb{R}^{2}$, we used
\begin{equation}
G_{\mathrm{sing}}^{2D}(\mathbf{x},\boldsymbol{\xi})
=
-\frac{1}{2\pi}
\ln \left( \|\mathbf{x}-\boldsymbol{\xi}\| \right),
\end{equation}
and decomposed the Green's function as
\begin{equation}
G(\mathbf{x},\boldsymbol{\xi})
=
G_{\mathrm{sing}}^{2D}(\mathbf{x},\boldsymbol{\xi})
+
G_{\mathrm{sm}}^{2D}(\mathbf{x},\boldsymbol{\xi}).
\end{equation}
This differs from the three-dimensional case, where the free-space singularity has the inverse-distance form $1/(4\pi\|\mathbf{x}-\boldsymbol{\xi}\|)$. Therefore, the singular component used in SDC-GON is dimension-dependent: the logarithmic singularity is used for the two-dimensional Poisson problem, whereas the inverse-distance singularity is used for the three-dimensional Poisson and reaction--diffusion benchmarks.

\subsection{SDC-GON architecture}

\begin{figure}[t]
\centering
\includegraphics[width=\linewidth]{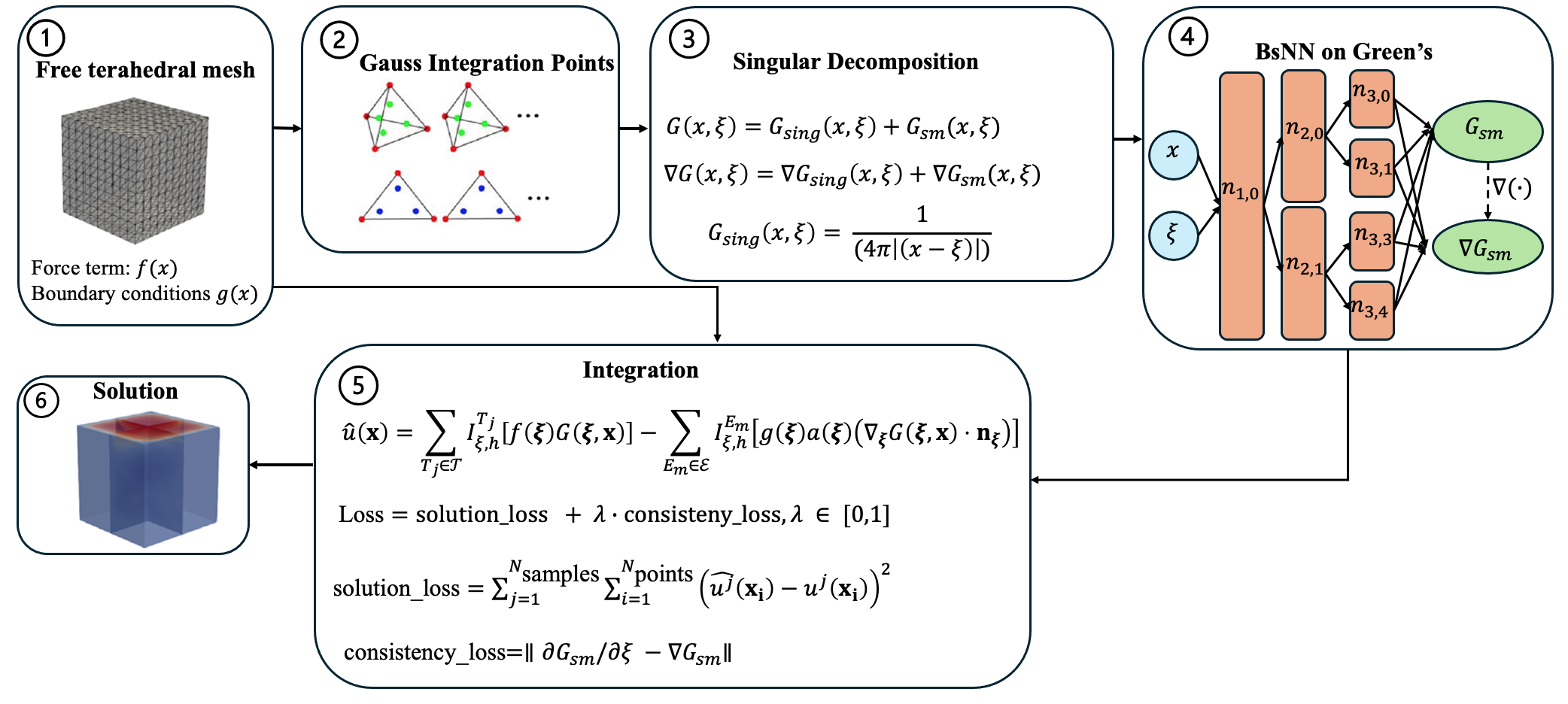}
\caption{Framework of SDC-GON. (1) Free tetrahedral mesh and physical conditions; (2) Gauss integration points; (3) singular decomposition; (4) shared BsNN backbone with dual output heads; (5) Volterra integration; (6) solution.}
\label{fig:sdc_gon_framework}
\end{figure}

The architecture of SDC-GON is shown in Fig.~\ref{fig:sdc_gon_framework}. The input to the network is the concatenated coordinate pair $(\mathbf{x},\boldsymbol{\xi})$, which forms a six dimensional vector in the three dimensional cases. A single shared BsNN backbone maps this input to a shared feature representation. Two output heads are attached to the backbone. The first head, denoted by $H_G(\mathbf{x},\boldsymbol{\xi};\boldsymbol{\theta})$, predicts the smooth correction $G_{\mathrm{sm}}(\mathbf{x},\boldsymbol{\xi})$ as a scalar. The second head, denoted by $\mathbf{H}_{\nabla}(\mathbf{x},\boldsymbol{\xi};\boldsymbol{\theta})$, predicts $\nabla_{\boldsymbol{\xi}}G_{\mathrm{sm}}(\mathbf{x},\boldsymbol{\xi})$ as a three dimensional vector.

\begin{figure}[t]
\centering
\includegraphics[width=\linewidth]{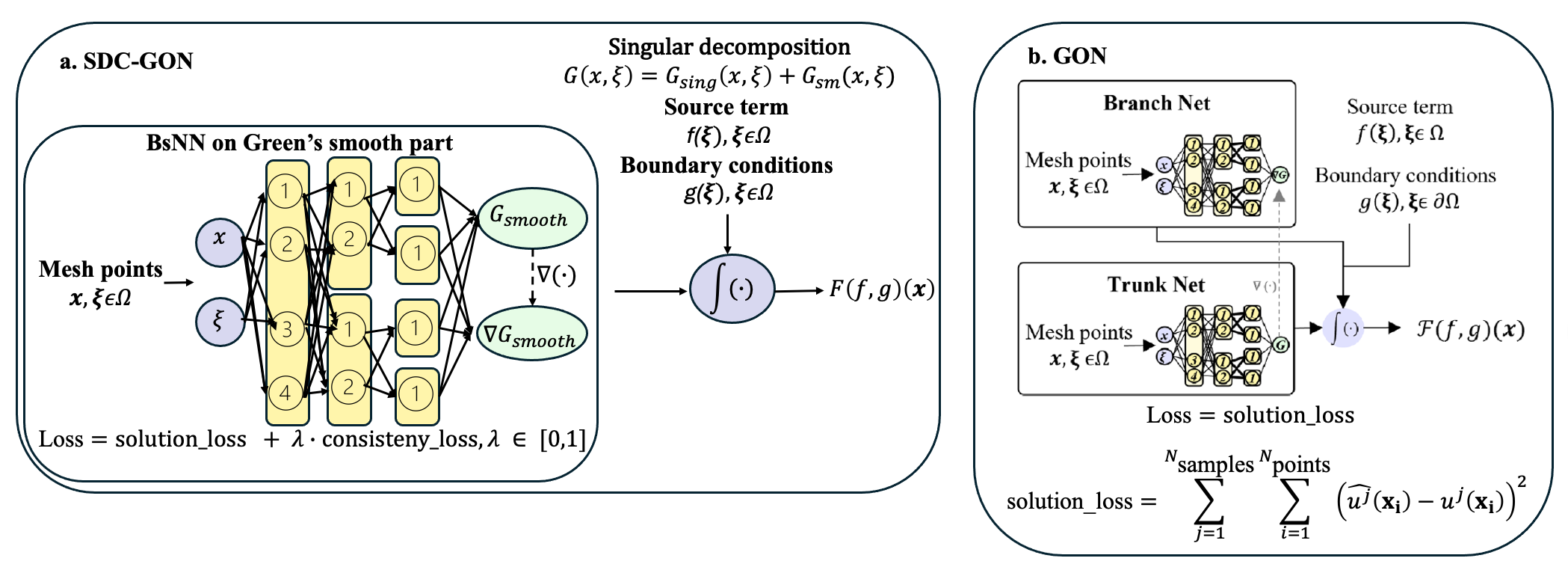}
\caption{Comparison of the original GON architecture and the proposed SDC-GON architecture.}
\label{fig:gon_sdc_comparison}
\end{figure}

The original GON uses two independent BsNN instances. The Trunk Net approximates $G(\mathbf{x},\boldsymbol{\xi})$, while the Branch Net approximates the auxiliary gradient $\nabla_{\boldsymbol{\xi}}G(\mathbf{x},\boldsymbol{\xi})$. Since these two networks have separate parameters, the learned scalar Green's function and the learned gradient are not required to share intermediate features or satisfy any explicit consistency relation. In contrast, SDC-GON uses a single shared BsNN backbone with two output heads to predict the smooth correction $G_{\mathrm{sm}}(\mathbf{x},\boldsymbol{\xi})$ and its source-coordinate gradient $\nabla_{\boldsymbol{\xi}}G_{\mathrm{sm}}(\mathbf{x},\boldsymbol{\xi})$. This design follows the principle of shared representation learning in multi-task neural networks, where related prediction tasks can benefit from common intermediate features when their outputs are generated from the same underlying structure \cite{caruana1997multitask,ruder2017overview}. In the present setting, the scalar correction and its gradient are not independent physical quantities; the gradient should be the derivative of the same smooth correction that enters the volume integral. The shared backbone therefore provides an architectural coupling between the two outputs, while the self consistency loss provides an explicit differential constraint between them. The architectural difference between GON and SDC-GON is illustrated in Fig.~\ref{fig:gon_sdc_comparison}.

The BsNN architecture itself is retained from GON. A BsNN is composed of neuron blocks arranged in a binary structured connectivity pattern. Except for the first and last hidden layers, each neuron block is connected to two neuron blocks in the next layer. Let $o_{i,j}$ denote the output of the $j$th neuron block in the $i$th hidden layer. Let $w_{i,j}$, $b_{i,j}$, and $\sigma_{i,j}$ denote the corresponding weight, bias, and activation function. The block propagation is written as
\begin{equation}
o_{i,j}
=
\sigma_{i-1,j}
\left(
w_{i-1,j}
o_{i-1,\lfloor j/2\rfloor}
+
b_{i-1,j}
\right),
\qquad
i=1,\ldots,n-1,\quad
j=0,\ldots,2^{i-1}-1.
\label{eq:bsnn_block}
\end{equation}
The outputs of the final hidden neuron blocks are concatenated into $o_{n-1}$, and the network output is obtained by
\begin{equation}
o_{n}
=
\sigma_{n-1}
\left(
w_{n-1}o_{n-1}
+
b_{n-1}
\right).
\label{eq:bsnn_output}
\end{equation}
This binary structure reduces dense coupling between all hidden neurons while preserving the ability to represent localized features. In SDC-GON, this structure is used to approximate the smooth correction rather than the raw singular Green's function.

\subsection{Integration with singular decomposition}

Substituting Eqs.~\eqref{eq:green_decomposition} and \eqref{eq:gradient_decomposition} into Eq.~\eqref{eq:green_representation} gives the decomposed solution representation
\begin{align}
u(\mathbf{x})
=
&
\int_{\Omega}
f(\boldsymbol{\xi})G_{\mathrm{sing}}(\mathbf{x},\boldsymbol{\xi})\,d\boldsymbol{\xi}
+
\int_{\Omega}
f(\boldsymbol{\xi})G_{\mathrm{sm}}(\mathbf{x},\boldsymbol{\xi})\,d\boldsymbol{\xi}
\nonumber\\
&
-
\int_{\partial\Omega}
g(\boldsymbol{\xi})a(\boldsymbol{\xi})
\left(
\nabla_{\boldsymbol{\xi}}G_{\mathrm{sing}}(\mathbf{x},\boldsymbol{\xi})\cdot \mathbf{n}_{\boldsymbol{\xi}}
\right)
\,dS(\boldsymbol{\xi})
\nonumber\\
&
-
\int_{\partial\Omega}
g(\boldsymbol{\xi})a(\boldsymbol{\xi})
\left(
\nabla_{\boldsymbol{\xi}}G_{\mathrm{sm}}(\mathbf{x},\boldsymbol{\xi})\cdot \mathbf{n}_{\boldsymbol{\xi}}
\right)
\,dS(\boldsymbol{\xi}).
\label{eq:decomposed_integral}
\end{align}
The volume integral involving $G_{\mathrm{sing}}$ uses the closed form singular kernel in Eq.~\eqref{eq:singular_green}. In the present implementation, the solution evaluation points $\mathbf{x}$ are restricted to mesh vertices. The volume integral involving $G_{\mathrm{sm}}$ is evaluated using standard Gaussian quadrature. The boundary integral involving $\nabla_{\boldsymbol{\xi}}G_{\mathrm{sing}}$ is computed from the analytical expression in Eq.~\eqref{eq:singular_gradient}. The boundary integral involving $\nabla_{\boldsymbol{\xi}}G_{\mathrm{sm}}$ uses the gradient head output $\mathbf{H}_{\nabla}(\mathbf{x},\boldsymbol{\xi};\boldsymbol{\theta})$ defined in the architecture description.

The same quadrature rules as GON are retained for the smooth integrals. A four point Gaussian quadrature rule is used on tetrahedral elements in the domain, and a three point Gaussian quadrature rule is used on triangular boundary faces. Let $\mathcal{T}=\{T_l\}$ denote the tetrahedral elements and let $\mathcal{E}_{\mathrm{bdry}}=\{E_m\}$ denote the triangular boundary faces. The numerical approximation of the decomposed solution can be written as
\begin{align}
\hat{u}(\mathbf{x})
\approx
&
\sum_{T_l\in\mathcal{T}}
I^{T_l}_{\boldsymbol{\xi},h}
\left[
f(\boldsymbol{\xi})
G_{\mathrm{sing}}(\mathbf{x},\boldsymbol{\xi})
\right]
+
\sum_{T_l\in\mathcal{T}}
I^{T_l}_{\boldsymbol{\xi},h}
\left[
f(\boldsymbol{\xi})
G_{\mathrm{sm}}(\mathbf{x},\boldsymbol{\xi})
\right]
\nonumber\\
&
-
\sum_{E_m\in\mathcal{E}_{\mathrm{bdry}}}
I^{E_m}_{\boldsymbol{\xi},h}
\left[
g(\boldsymbol{\xi})a(\boldsymbol{\xi})
\left(
\nabla_{\boldsymbol{\xi}}G_{\mathrm{sing}}(\mathbf{x},\boldsymbol{\xi})
\cdot
\mathbf{n}_{\boldsymbol{\xi}}
\right)
\right]
\nonumber\\
&
-
\sum_{E_m\in\mathcal{E}_{\mathrm{bdry}}}
I^{E_m}_{\boldsymbol{\xi},h}
\left[
g(\boldsymbol{\xi})a(\boldsymbol{\xi})
\left(
\mathbf{H}_{\nabla}(\mathbf{x},\boldsymbol{\xi};\boldsymbol{\theta})
\cdot
\mathbf{n}_{\boldsymbol{\xi}}
\right)
\right],
\label{eq:quadrature_decomposed}
\end{align}
where $I^{T_l}_{\boldsymbol{\xi},h}$ and $I^{E_m}_{\boldsymbol{\xi},h}$ denote the quadrature operators on tetrahedral elements and boundary faces, respectively. The term $\mathbf{H}_{\nabla}(\mathbf{x},\boldsymbol{\xi};\boldsymbol{\theta})$ denotes the three dimensional output of the gradient head introduced in the architecture description. Since the learned integrands involve only the smooth correction and its smooth gradient, the standard quadrature rules are applied to well behaved neural outputs rather than to the full singular Green's function.

\subsection{Self consistency loss}

The training objective contains a solution reconstruction loss and a self consistency loss. Let $\hat{u}^{j}(\mathbf{x}_{i})$ be the solution predicted by the decomposed integral formula for the $j$th training sample at the evaluation point $\mathbf{x}_{i}$, and let $u^{j}(\mathbf{x}_{i})$ be the reference FEM solution. The solution loss is
\begin{equation}
\mathcal{L}_{\mathrm{sol}}
=\sum_{j=1}^{N_{\mathrm{samples}}}
\sum_{i=1}^{N_{\mathrm{points}}}
\left(
\hat{u}^{j}(\mathbf{x}_{i})
-
u^{j}(\mathbf{x}_{i})
\right)^{2}.
\label{eq:solution_loss}
\end{equation}

Let $H_{G}(\mathbf{x},\boldsymbol{\xi};\boldsymbol{\theta})$ denote the scalar output head that predicts $G_{\mathrm{sm}}(\mathbf{x},\boldsymbol{\xi})$, and let $\mathbf{H}_{\nabla}(\mathbf{x},\boldsymbol{\xi};\boldsymbol{\theta})$ denote the vector output head that predicts $\nabla_{\boldsymbol{\xi}}G_{\mathrm{sm}}(\mathbf{x},\boldsymbol{\xi})$. The consistency loss is defined as
\begin{equation}
\mathcal{L}_{\mathrm{con}}
=\sum_{k=1}^{N_{\mathrm{c}}}
\left\|
\nabla_{\boldsymbol{\xi}}
H_{G}(\mathbf{x}_{k},\boldsymbol{\xi}_{k};\boldsymbol{\theta})
-
\mathbf{H}_{\nabla}(\mathbf{x}_{k},\boldsymbol{\xi}_{k};\boldsymbol{\theta})
\right\|_{2}^{2}.
\label{eq:consistency_loss}
\end{equation}
In the implementation, the pairs $(\mathbf{x}_{k},\boldsymbol{\xi}_{k})$ are sampled from the same coordinate pairs used in the quadrature mini-batch for the solution loss. Since the consistency term is imposed on the smooth correction $G_{\mathrm{sm}}$ rather than on the raw singular Green's function, no exclusion of near diagonal pairs is required. 

The total loss at epoch $e$ is
\begin{equation}
\mathcal{L}_{\mathrm{total}}(e)
=
\mathcal{L}_{\mathrm{sol}}
+
\lambda(e)\mathcal{L}_{\mathrm{con}},
\label{eq:total_loss}
\end{equation}
with
\begin{equation}
\lambda(e)
=
\begin{cases}
0, & e<E_{\mathrm{c}},\\
\lambda_{\mathrm{c}}, & e\ge E_{\mathrm{c}}.
\end{cases}
\label{eq:lambda_schedule}
\end{equation}
Here $E_{\mathrm{c}}$ is the activation epoch for the consistency term, and $\lambda_{\mathrm{c}}$ is the consistency coefficient used after activation.

The consistency loss is applied only to the smooth correction. After singular decomposition, the learned scalar output represents a smooth correction, so its automatic differentiation gradient is a reliable target for the learned gradient head. The consistency loss therefore penalizes genuine disagreement between two smooth representations. The consistency term is activated after an initial warmup period, during which only the solution loss is minimized. This schedule allows the smooth Green's correction to become sufficiently accurate before the gradient consistency signal is introduced. 

\subsection{Warmup cosine learning rate schedule}

The optimization of SDC-GON involves learning a smooth Green's function correction and its source coordinate gradient through a coupled solution reconstruction and consistency regularization objective. Since the two loss components may contribute gradients with different magnitudes during different stages of training, a learning rate schedule is used to control the step size throughout optimization. SDC-GON adopts a warmup cosine schedule in which the learning rate is kept at the base value during the initial training stage and is then smoothly reduced toward a prescribed minimum value. The initial constant stage allows the network to learn the dominant solution structure efficiently, while the later decay stage reduces parameter oscillation and supports stable refinement of the learned smooth correction and gradient head.

Let $\eta(e)$ denote the learning rate at epoch $e$, $\eta_{0}$ denote the base learning rate, $\eta_{\min}$ denote the minimum learning rate, $E_{\mathrm{w}}$ denote the warmup epoch, and $E_{\max}$ denote the total number of training epochs. The warmup cosine schedule is defined as
\begin{equation}
\eta(e)
=
\begin{cases}
\eta_{0},
&
0\le e<E_{\mathrm{w}},\\
\eta_{\min}
+
\frac{1}{2}
\left(
\eta_{0}-\eta_{\min}
\right)
\left[
1+
\cos
\left(
\pi
\frac{e-E_{\mathrm{w}}}{E_{\max}-E_{\mathrm{w}}}
\right)
\right],
&
E_{\mathrm{w}}\le e\le E_{\max}.
\end{cases}
\label{eq:lr_schedule}
\end{equation}
The schedule is independent of the benchmark geometry and is applied to all scalar and vector valued cases considered in this work. 

\subsection{Methodological comparison with Green's function learning methods}

Table~\ref{tab:method_comparison} summarizes the methodological characteristics of representative Green's function learning methods and the proposed SDC-GON framework. The comparison covers the neural backbone, loss formulation, integration strategy, boundary condition handling, source term handling, problem dimensionality, singularity treatment, and consistency enforcement. Earlier Green's function learning methods demonstrate different ways of approximating Green's functions, including rational neural networks, feedforward neural networks with PDE constraints, and Monte Carlo or Gaussian integration strategies. GON extends this direction to three dimensional bounded domains and supports varying boundary conditions and source terms through a data constrained Green's function integral formulation. SDC-GON preserves the boundary invariant and source term invariant formulation of GON while introducing two structural changes that are absent from the other compared methods: the singular part is treated analytically so that the neural network learns only the smooth correction, and the learned smooth correction is regularized to be consistent with its predicted source coordinate gradient.

\begin{table}[t]
\centering
\caption{Methodological comparison of representative Green's function learning methods and SDC-GON. In the Singularity column, ``Indirect'' indicates that no known singular component is analytically subtracted; instead the singularity is handled implicitly through architectural flexibility (rational NN, BsNN) or smooth-surrogate training. ``Analytical'' indicates explicit subtraction of the free-space singular kernel before neural approximation.}
\label{tab:method_comparison}
\small
\begin{tabular}{lllllllll}
\hline
Method & Backbone & Loss & Integration & BC & Source & Dim. & Singularity & Consistency \\
\hline
Boulle et al. \cite{boulle2022data} & Rational NN & Data & Monte Carlo & No & Yes & 1D--2D & Indirect & No \\
Teng et al. \cite{teng2022learning} & FNN & PDE & Gauss & Yes & Yes & 1D--2D & Indirect & No \\
Aldirany et al. \cite{aldirany2024operator} & FNN & PDE & Monte Carlo & No & Yes & 1D--2D & Indirect & No \\
GON \cite{gu2025explainable} & BsNN & Data & Gauss & Yes & Yes & 3D & Indirect & No \\
SDC-GON & Shared BsNN & Data + consistency & Gauss & Yes & Yes & 3D & Analytical & Yes \\
\hline
\end{tabular}
\end{table}

The methods of Boulle et al. \cite{boulle2022data}, Teng et al. \cite{teng2022learning}, and Aldirany et al. \cite{aldirany2024operator} provide important precedents for learning Green's functions, but their reported settings are limited to lower dimensional problems or fixed boundary condition treatments. GON extends Green's function learning to three dimensional bounded domains and supports varying boundary conditions and source terms. SDC-GON builds on this capability and targets the remaining structural limitations. The singular part is handled analytically, the neural network learns only the smooth correction, and the learned scalar correction is made consistent with the learned gradient through explicit regularization.

\section{Experiments and Results}
\label{results}

This section evaluates SDC-GON on five benchmark cases: the two dimensional Poisson equation, three dimensional steady heat conduction, heterogeneous reaction diffusion on a flat plate, heterogeneous reaction diffusion in a pipe, and three dimensional Stokes flow. The two dimensional Poisson case is used to compare SDC-GON with existing Green's function learning methods in a setting commonly considered in prior studies. The remaining four cases evaluate whether the proposed singular decomposition and consistency regularization remain effective for three dimensional bounded domains, irregular geometries, heterogeneous coefficients, varying boundary conditions, varying source terms, and vector valued systems.

All datasets were generated using COMSOL Multiphysics. The compared baselines include PINN \cite{raissi2019physics}, DeepONet \cite{lu2021learning}, PI-DeepONet \cite{wang2021learning}, FNO \cite{li2021fourier}, and GON \cite{gu2025explainable}. The SDC-GON experiments were conducted on an NVIDIA GeForce RTX 3090 GPU with 24 GB memory. Because of this memory constraint, the SDC-GON models for the pipe and Stokes cases use smaller BsNN layer widths than the larger GON reference configurations. This setting provides a stricter comparison, since SDC-GON is evaluated with a smaller architecture while being compared against larger GON models in these two cases.

\subsection{Experimental settings and evaluation metric}

The same SDC-GON optimization settings were used across all experiments to avoid benchmark-specific hyperparameter tuning. The consistency term was activated at epoch $E_{\mathrm{c}}=3500$ with coefficient $\lambda_{\mathrm{c}}=2\times10^{-5}$. The warmup cosine learning rate schedule used base learning rate $\eta_{0}=10^{-3}$, minimum learning rate $\eta_{\min}=10^{-5}$, warmup length $E_{\mathrm{w}}=500$, and total training length $E_{\max}=10000$.

All training and test errors reported in this section were computed using the mean squared error (MSE). The same MSE definition was used for SDC-GON and all baseline models. Each experiment was independently repeated three times using different random initializations and data-shuffling orders. The values reported in the tables are the mean MSE over the three independent runs. For PINN, the training error is not reported in the result tables because PINN is trained on a per-instance basis by minimizing a physics residual objective; it does not produce a unified aggregate training loss over multiple solution instances and therefore does not yield a training error in the same sense as the data-driven operator learning baselines.

The hyperparameters used in SDC-GON were selected based on preliminary training-log observations and then fixed across all benchmarks. The consistency loss was not activated from the beginning of training because, during early epochs, the solution-fitting loss and the learned smooth correction were still changing rapidly. Applying the consistency constraint too early could force agreement between unstable estimates of the smooth correction and its gradient. Therefore, the consistency term was activated only after the main training loss entered a relatively stable decreasing regime, which led to the choice of $E_{\mathrm{c}}=3500$.

The consistency-loss coefficient was chosen conservatively. The value $\lambda_{\mathrm{c}}=2\times10^{-5}$ was selected so that the consistency term regularized the learned correction and gradient without dominating the primary solution-fitting objective. Preliminary training logs showed that substantially larger consistency weights could interfere with fitting the reference Green's function values, whereas very small weights had limited effect. The learning-rate schedule was selected using the same practical criterion: $\eta_{0}=10^{-3}$ was used as a stable initial learning rate, the warmup stage with $E_{\mathrm{w}}=500$ was introduced to reduce unstable early updates, and cosine decay to $\eta_{\min}=10^{-5}$ was used for late-stage refinement.

\subsection{Two dimensional Poisson equation}

\begin{figure}[t]
\centering
\includegraphics[width=\linewidth]{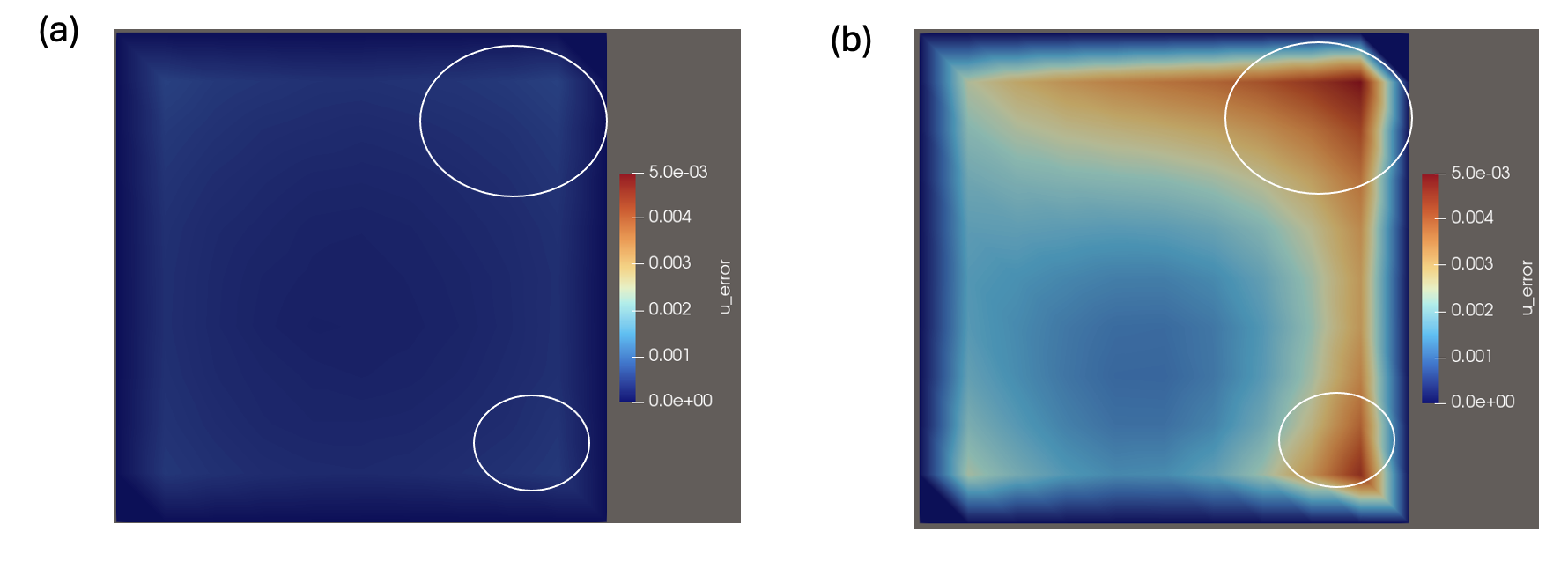}
\caption{Two dimensional Poisson case: absolute pointwise error for SDC-GON (a) and GON (b). Both panels use the same error color scale, and the circled regions indicate areas where the visual difference between the two methods is most evident.}
\label{fig:poisson2d}
\end{figure}

The two dimensional Poisson case is included to compare SDC-GON with existing Green's function based methods. It considers a square domain with homogeneous Dirichlet boundary conditions and sinusoidal exact solutions. Table~\ref{tab:poisson2d} reports the training and testing errors for Boulle et al. \cite{boulle2022data}, Teng et al. \cite{teng2022learning}, Aldirany et al. \cite{aldirany2024operator}, GON \cite{gu2025explainable}, and SDC-GON. Figure~\ref{fig:poisson2d} shows the pointwise error comparison between SDC-GON and GON, which is the strongest baseline in Table~\ref{tab:poisson2d}.

\begin{table}[t]
\centering
\caption{Comparison with existing Green's function based methods for the two dimensional Poisson case.}
\label{tab:poisson2d}
\begin{tabular}{lll}
\hline
Method & Training error & Testing error \\
\hline
Boulle et al. \cite{boulle2022data} & $2.51\times 10^{-3}$ & $2.56\times 10^{-3}$ \\
Teng et al. \cite{teng2022learning} & $2.63\times 10^{-2}$ & $1.05\times 10^{-1}$ \\
Aldirany et al. \cite{aldirany2024operator} & $9.75\times 10^{-3}$ & $3.91\times 10^{-2}$ \\
GON \cite{gu2025explainable} & $2.32\times 10^{-5}$ & $3.14\times 10^{-5}$ \\
SDC-GON & $\mathbf{4.86\times 10^{-8}}$ & $\mathbf{8.64\times 10^{-8}}$ \\
\hline
\end{tabular}
\end{table}

As shown in Table~\ref{tab:poisson2d}, SDC-GON gives the lowest training and testing errors among the compared Green's function based methods. Its testing error is $8.64\times 10^{-8}$, compared with $3.14\times 10^{-5}$ for GON. This corresponds to an improvement factor of approximately $360$ relative to GON on this benchmark. These results establish the accuracy advantage of SDC-GON on the two dimensional Poisson benchmark before the method is evaluated on the three dimensional cases.

The particularly large improvement observed for the two-dimensional Poisson benchmark is consistent with the structure of the Green's function singularity. In two dimensions, the dominant local singular behavior of the Poisson Green's function is logarithmic. By explicitly subtracting the term $-(1/2\pi)\ln\|\mathbf{x}-\boldsymbol{\xi}\|$, the remaining correction $G_{\mathrm{sm}}^{2D}$ becomes substantially smoother near $\mathbf{x}=\boldsymbol{\xi}$ than the original Green's function. Consequently, the neural network is not required to approximate the non-smooth logarithmic singularity directly, which greatly reduces the learning difficulty. However, the correction term is not identically trivial, because it still encodes the effects of the bounded square domain and the imposed boundary conditions. Therefore, the large error reduction in the two-dimensional Poisson case is a natural consequence of using the correct dimension-dependent singular decomposition, rather than an unexplained empirical effect.

Figure~\ref{fig:poisson2d} provides the corresponding pointwise error comparison. The SDC-GON error field in Fig.~\ref{fig:poisson2d}(a) remains close to zero over nearly the entire square domain, including the two circled regions. In contrast, the GON error field in Fig.~\ref{fig:poisson2d}(b) shows visibly larger errors concentrated near the upper right corner, the lower right region, and along the upper and right boundary regions. Both plots use the same error color scale from $0$ to $5.0\times 10^{-3}$, so the visual comparison is consistent with the testing error reduction reported in Table~\ref{tab:poisson2d}. The figure indicates that the improvement is not limited to the global error metric, but is also reflected in the spatial distribution of the pointwise error.

\subsection{Three dimensional steady heat conduction}

\begin{figure}[t]
\centering
\includegraphics[width=\linewidth]{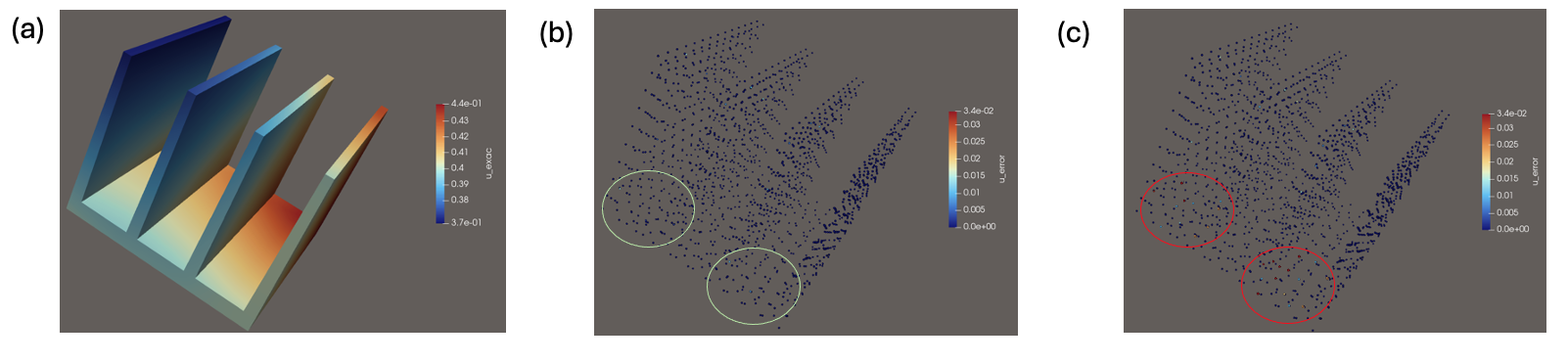}
\caption{Three dimensional heat conduction case: exact solution field (a), pointwise error for SDC-GON (b), and pointwise error for GON (c) on the finned tube geometry. Highlighted regions indicate areas of visible error reduction.}
\label{fig:finned_tube}
\end{figure}

The second benchmark considers steady heat conduction on a finned tube geometry with varying Gaussian random field boundary conditions. The mesh contains 4427 tetrahedral cells, 1478 vertices, and 2968 triangular boundary faces. A total of 100 samples were generated, with 70 samples used for training and 30 samples used for testing. The finned tube geometry is irregular, which makes this benchmark more challenging for methods that rely on regular grids or interpolation between structured and unstructured representations. Table~\ref{tab:finned_tube} reports the layer configurations and errors. Figure~\ref{fig:finned_tube} shows the exact solution and the pointwise error comparison between SDC-GON and GON.

\begin{table}[t]
\centering
\caption{Performance of different baseline models on the three dimensional heat conduction case on the finned tube geometry.}
\label{tab:finned_tube}
\small
\begin{tabular}{llll}
\hline
Model & Layers & Training error & Testing error \\
\hline
PINN \cite{raissi2019physics} & $[3,12,12,12,1]$ & not reported & $2.56\times 10^{-2}$ \\
PI-DeepONet \cite{wang2021learning} & Trunk $[3,12,12,12,1]$ & $4.14\times 10^{-4}$ & $1.78\times 10^{-3}$ \\
DeepONet \cite{lu2021learning} & Trunk $[3,12,12,12,1]$ & $8.14\times 10^{-4}$ & $1.37\times 10^{-3}$ \\
FNO \cite{li2021fourier} & $[2,3,3,3,3,3,1]$ & $1.01\times 10^{-4}$ & $6.32\times 10^{-4}$ \\
GON \cite{gu2025explainable} & $[6,12,12,12,1]$ & $1.50\times 10^{-5}$ & $5.00\times 10^{-5}$ \\
SDC-GON & $[6,12,12,12,1]$ & $\mathbf{7.30\times 10^{-6}}$ & $\mathbf{8.20\times 10^{-6}}$ \\
\hline
\end{tabular}
\end{table}

The proposed method achieves the best performance on the finned tube case. As shown in Table~\ref{tab:finned_tube}, SDC-GON obtains a testing error of $8.20\times 10^{-6}$, which is lower than $5.00\times 10^{-5}$ for GON, $6.32\times 10^{-4}$ for FNO, $1.37\times 10^{-3}$ for DeepONet, $1.78\times 10^{-3}$ for PI-DeepONet, and $2.56\times 10^{-2}$ for PINN. Since both GON and SDC-GON use the same layer width $[6,12,12,12,1]$ in this case, the reduction in testing error is attributed to the structural changes in SDC-GON rather than an increase in model size.

As shown in Fig.~\ref{fig:finned_tube}, the visual error fields are consistent with the quantitative results in Table~\ref{tab:finned_tube}. The exact solution in Fig.~\ref{fig:finned_tube}(a) shows the temperature distribution on the finned tube geometry, where the irregular fins create a nontrivial spatial solution pattern. The SDC-GON error field in Fig.~\ref{fig:finned_tube}(b) remains sparse and low over most of the geometry, with only small localized errors in the highlighted regions. In contrast, the GON error field in Fig.~\ref{fig:finned_tube}(c) contains more visible localized error points in the same highlighted regions. Both error plots use the same color scale from $0$ to $3.4\times 10^{-2}$, so the lower visible error concentration for SDC-GON agrees with the testing error reduction for SDC-GON.

\subsection{Heterogeneous reaction diffusion equations}

\begin{figure}[t]
\centering
\includegraphics[width=\linewidth]{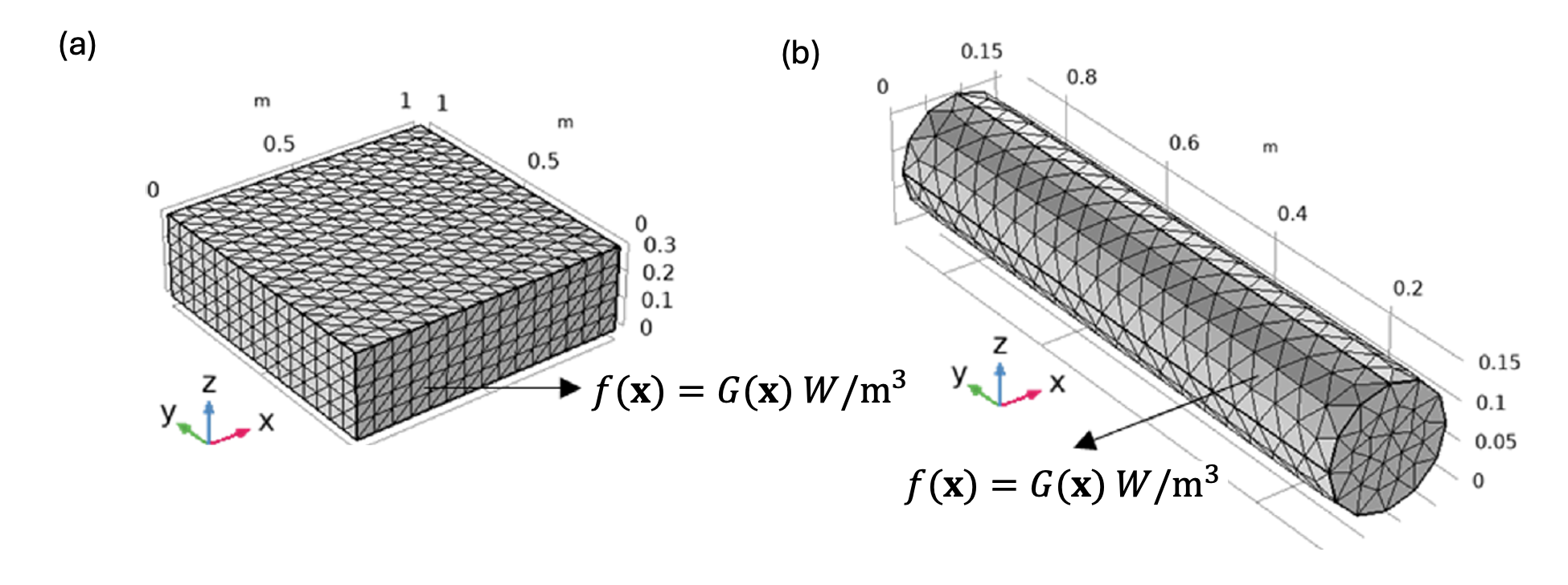}
\caption{Computational meshes for the flat plate geometry (a) and the pipe geometry (b). Both reaction diffusion cases use a Gaussian random field source term $f(\mathbf{x})=G(\mathbf{x})~\mathrm{W}/\mathrm{m}^{3}$.}
\label{fig:reaction_setup}
\end{figure}

The third and fourth benchmarks evaluate heterogeneous reaction diffusion equations. The flat plate case uses a smoothly varying diffusion coefficient, while the pipe case uses a discontinuous step function diffusion coefficient representing a material interface in a microfluidic channel. Figure~\ref{fig:reaction_setup} shows the computational meshes for these two cases. These benchmarks test two different types of heterogeneity: smooth coefficient variation in the plate and discontinuous coefficient variation in the pipe.

\begin{figure}[t]
\centering
\includegraphics[width=\linewidth]{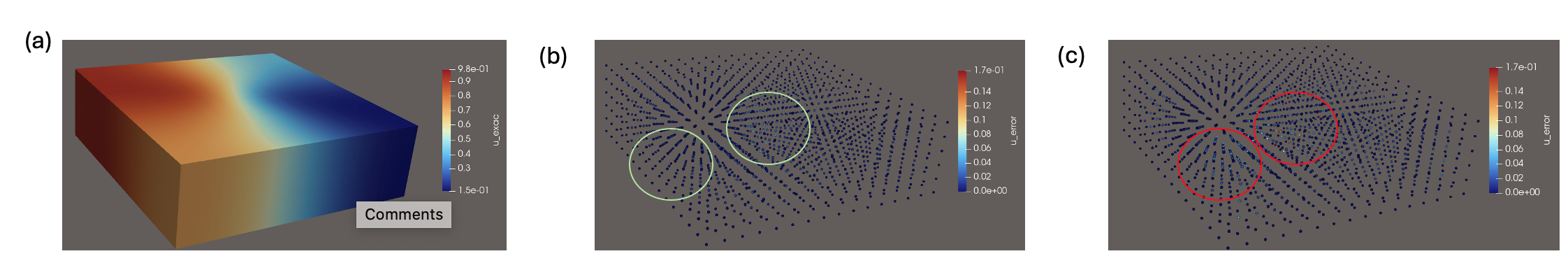}
\caption{Flat plate case: exact solution field (a), pointwise error for SDC-GON (b), and pointwise error for GON (c). Highlighted regions show localized error differences between the two methods.}
\label{fig:plate_error}
\end{figure}

For the flat plate case, the diffusion coefficient varies smoothly as $a(\mathbf{x})=(x-0.5)^2+(y-0.5)^2+(z-0.1)^2$. The dataset uses 6750 tetrahedral cells, 1536 vertices, and 1500 triangular boundary faces. A total of 100 samples were generated, with 70 samples used for training and 30 samples used for testing. Table~\ref{tab:plate} reports the quantitative results, and Fig.~\ref{fig:plate_error} presents the pointwise error comparison.

\begin{table}[t]
\centering
\caption{Performance of different baseline models on the flat plate reaction diffusion case.}
\label{tab:plate}
\small
\begin{tabular}{llll}
\hline
Model & Layers & Training error & Testing error \\
\hline
PINN \cite{raissi2019physics} & $[3,12,12,12,1]$ & not reported & $6.54\times 10^{-2}$ \\
PI-DeepONet \cite{wang2021learning} & Trunk $[3,12,12,12,1]$ & $4.02\times 10^{-2}$ & $3.93\times 10^{-2}$ \\
DeepONet \cite{lu2021learning} & Trunk $[3,12,12,12,1]$ & $3.68\times 10^{-2}$ & $3.45\times 10^{-2}$ \\
FNO \cite{li2021fourier} & $[2,6,6,6,6,1]$ & $1.33\times 10^{-3}$ & $2.89\times 10^{-3}$ \\
GON \cite{gu2025explainable} & $[6,12,12,12,1]$ & $3.80\times 10^{-4}$ & $2.52\times 10^{-3}$ \\
SDC-GON & $[6,12,12,12,1]$ & $\mathbf{1.00\times 10^{-4}}$ & $\mathbf{2.78\times 10^{-4}}$ \\
\hline
\end{tabular}
\end{table}

On the flat plate, SDC-GON achieves a testing error of $2.78\times 10^{-4}$, which is lower than $2.52\times 10^{-3}$ for GON and $2.89\times 10^{-3}$ for FNO, as shown in Table~\ref{tab:plate}. The reduction relative to GON is obtained with the same layer width $[6,12,12,12,1]$. Compared with DeepONet and PI-DeepONet, whose testing errors are $3.45\times 10^{-2}$ and $3.93\times 10^{-2}$, the Green's function integral formulation gives a much lower error on this heterogeneous scalar problem. The flat plate case shows the largest relative improvement among the three dimensional scalar cases, with an approximately nine fold testing error reduction relative to GON. This suggests that the proposed decomposition is especially effective when the coefficient field varies smoothly and the remaining correction can be represented more regularly by the neural network.

Figure~\ref{fig:plate_error} provides the corresponding spatial error comparison. The exact solution in Fig.~\ref{fig:plate_error}(a) shows a smooth transition across the plate, with the strongest solution variation occurring from the high value region on the left to the low value region on the right. The SDC-GON error field in Fig.~\ref{fig:plate_error}(b) shows localized residual errors mainly in the transition region where the solution gradient is steepest. The GON error field in Fig.~\ref{fig:plate_error}(c) shows more visible error concentrations in the highlighted areas. Both error plots use the same color scale from $0$ to $1.7\times 10^{-1}$, so the visual reduction in pointwise error is consistent with the testing error decrease from $2.52\times 10^{-3}$ for GON to $2.78\times 10^{-4}$ for SDC-GON.

\begin{figure}[t]
\centering
\includegraphics[width=\linewidth]{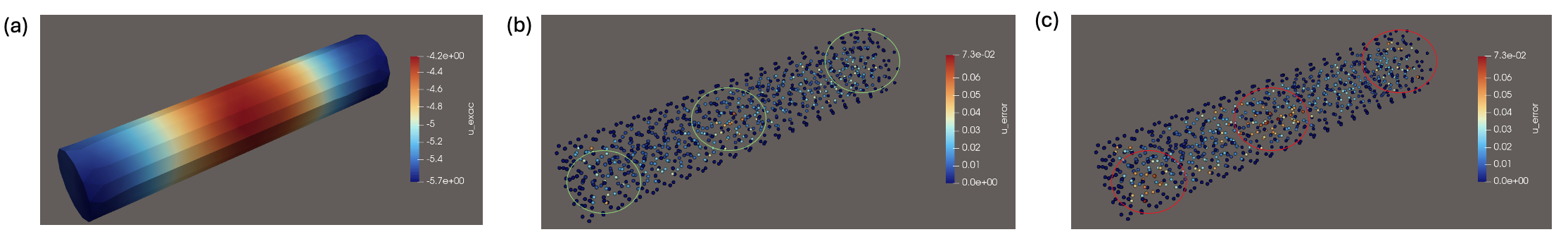}
\caption{Pipe case: exact solution field (a), pointwise error for SDC-GON (b), and pointwise error for GON (c). SDC-GON uses the smaller BsNN architecture, while GON is shown for comparison using the same error scale.}
\label{fig:pipe_error}
\end{figure}

For the pipe case, the diffusion coefficient is defined as $a(\mathbf{x})=0.002$ for $x<0.08$ m and $a(\mathbf{x})=0.001$ for $x\geq 0.08$ m, representing a discontinuous material interface between two substances. The pipe has a radius of 0.08 m and a length of 0.8 m. The mesh contains 3904 tetrahedral cells, 868 vertices, and 740 triangular boundary faces. A total of 100 samples were generated, with 70 samples used for training and 30 samples used for testing. Table~\ref{tab:pipe} reports the quantitative results, and Fig.~\ref{fig:pipe_error} shows the pointwise error comparison.

\begin{table}[t]
\centering
\caption{Performance of different baseline models on the pipe reaction diffusion case.}
\label{tab:pipe}
\small
\begin{tabular}{llll}
\hline
Model & Layers & Training error & Testing error \\
\hline
PINN \cite{raissi2019physics} & $[3,24,24,24,1]$ & not reported & $1.64$ \\
PI-DeepONet \cite{wang2021learning} & Trunk $[3,24,24,24,1]$ & $4.72\times 10^{-3}$ & $4.63\times 10^{-2}$ \\
DeepONet \cite{lu2021learning} & Trunk $[3,24,24,24,1]$ & $1.95\times 10^{-3}$ & $2.84\times 10^{-2}$ \\
FNO \cite{li2021fourier} & $[2,12,12,12,12,1]$ & $1.72\times 10^{-3}$ & $2.14\times 10^{-3}$ \\
GON small \cite{gu2025explainable} & $[6,12,12,12,1]$ & $2.20\times 10^{-4}$ & $9.60\times 10^{-4}$ \\
GON large \cite{gu2025explainable} & $[6,24,24,24,1]$ & $4.43\times 10^{-4}$ & $4.63\times 10^{-4}$ \\
SDC-GON & $[6,12,12,12,1]$ & $\mathbf{3.00\times 10^{-4}}$ & $\mathbf{3.70\times 10^{-4}}$ \\
\hline
\end{tabular}
\end{table}

The pipe case provides the clearest evidence that structural improvements can be more effective than simply increasing network width. Due to GPU memory constraints on the RTX 3090, SDC-GON uses the smaller BsNN architecture $[6,12,12,12,1]$ in Table~\ref{tab:pipe}. With this smaller architecture, SDC-GON achieves a testing error of $3.70\times 10^{-4}$. This is lower than the $9.60\times 10^{-4}$ testing error of GON with the same smaller width, and it is also lower than the $4.63\times 10^{-4}$ testing error of the larger GON configuration $[6,24,24,24,1]$. It is also notable that GON small achieves a lower training error ($2.20\times10^{-4}$) than SDC-GON ($3.00\times10^{-4}$) yet a substantially higher testing error ($9.60\times10^{-4}$ versus $3.70\times10^{-4}$), a train-to-test gap of approximately $4.4\times$ that suggests the independent dual-network design of GON small overfits the training samples more readily than the shared-backbone architecture of SDC-GON, which imposes an implicit coupling between the scalar Green's function and its gradient. The testing error is the more relevant comparison for generalization, and the testing error values show that singular decomposition and consistency regularization provide a clear gain on the discontinuous coefficient benchmark.

As discussed in the methodology section, the singular term used in the present framework is the constant-coefficient free-space singularity rather than the locally coefficient-adapted singularity. Therefore, for the heterogeneous pipe benchmark, where $a(\mathbf{x})$ is piecewise constant, the smooth correction may still contain a residual coefficient-dependent singular component. The improved performance of SDC-GON in this case indicates that the learned correction can effectively compensate for this residual in practice. Nevertheless, using the locally adapted singular term $1/(4\pi a(\boldsymbol{\xi})\|\mathbf{x}-\boldsymbol{\xi}\|)$ may further reduce the approximation burden in high-contrast heterogeneous media and will be considered in future work.

Figure~\ref{fig:pipe_error} shows the spatial distribution of the error for the pipe case. The exact solution in Fig.~\ref{fig:pipe_error}(a) varies along the length of the pipe, with a high magnitude region in the central portion and lower values near the two ends. The SDC-GON error field in Fig.~\ref{fig:pipe_error}(b) shows three highlighted regions with localized residual errors. In contrast, the GON error field in Fig.~\ref{fig:pipe_error}(c) shows stronger error concentrations in the corresponding highlighted regions. The GON panel contains more visible yellow and red points than the SDC-GON panel, especially around the left portion, middle transition region, and right end of the pipe. Both error fields use the same color scale from $0$ to $7.3\times 10^{-2}$, so the visual comparison is consistent with the testing error reduction for SDC-GON. This spatial error pattern indicates that the proposed structure reduces both the magnitude and the spatial extent of the pointwise error in the discontinuous coefficient case.

\subsection{Three-dimensional Stokes equations}

The final benchmark considers the three-dimensional lid-driven cavity Stokes problem. The governing system involves the velocity vector and pressure field. The body force is generated from a Gaussian random field, and the cavity uses no-slip boundary conditions on all walls except the moving top wall. The mesh contains 1331 vertices, 6000 tetrahedral cells, and 1200 triangular boundary faces. A total of 100 samples were generated, with 70 samples used for training and 30 samples used for testing.

The Stokes benchmark was handled separately from the scalar Poisson and diffusion--reaction benchmarks. For the scalar cases, the singular decomposition is based on the scalar free-space Green's function. In contrast, the three-dimensional Stokes Green's function is matrix-valued and has a tensorial singular structure. In free space, the Stokeslet has the form
\begin{equation}
G_{ij}^{\mathrm{Stokes}}(\mathbf{x},\boldsymbol{\xi})
=
\frac{1}{8\pi\mu}
\left(
\frac{\delta_{ij}}{\|\mathbf{x}-\boldsymbol{\xi}\|}
+
\frac{(x_i-\xi_i)(x_j-\xi_j)}
{\|\mathbf{x}-\boldsymbol{\xi}\|^{3}}
\right),
\end{equation}
where $\mu$ is the dynamic viscosity and $\delta_{ij}$ is the Kronecker delta. Therefore, the scalar singular component $G_{\mathrm{sing}}=1/(4\pi\|\mathbf{x}-\boldsymbol{\xi}\|)$ used for the diffusion-type problems is not used as an analytical singular correction for the Stokes benchmark.

Instead, the Stokes case is treated as a matrix-valued Green's function learning problem. The Green's function is a $3\times3$ matrix that maps force components to velocity components, and each row of the matrix is learned independently by a separate pair of dual-head networks in SDC-GON. For the boundary contribution, the Stokes benchmark uses a separate computational branch in which the normal derivative is computed by automatic differentiation from the learned Green's function. The analytical normal derivative of the scalar diffusion singular kernel is not added to the Stokes boundary term.

This treatment should be distinguished from a full Stokes boundary integral equation formulation, where the boundary terms involve the Stokes stress tensor and double-layer potential operators. The present Stokes experiment is therefore intended to evaluate whether the proposed learning architecture can approximate matrix-valued Green's function components under a row-wise learning strategy, rather than to derive a complete stress-based Stokes boundary integral solver. Developing a Stokes-specific singular decomposition based on the Stokeslet and its associated traction kernels is an important extension of the present framework.

\begin{table}[t]
\centering
\caption{Performance of different baseline models on the Stokes cavity case.}
\label{tab:stokes}
\small
\begin{tabular}{llll}
\hline
Model & Layers & Training error & Testing error \\
\hline
PINN \cite{raissi2019physics} & $[3,12,12,12,1]$ & not reported & $1.08\times 10^{-2}$ \\
PI-DeepONet \cite{wang2021learning} & Trunk $[3,12,24,12,4]$ & $5.59\times 10^{-3}$ & $5.31\times 10^{-3}$ \\
DeepONet \cite{lu2021learning} & Trunk $[3,12,24,12,4]$ & $2.15\times 10^{-3}$ & $2.17\times 10^{-3}$ \\
FNO \cite{li2021fourier} & $[2,12,12,12,12,3]$ & $1.36\times 10^{-3}$ & $2.21\times 10^{-3}$ \\
GON small \cite{gu2025explainable} & $[6,4,6,8,1]$ & $2.15\times 10^{-3}$ & $1.97\times 10^{-3}$ \\
GON large \cite{gu2025explainable} & $[6,12,24,12,1]$ & $5.71\times 10^{-4}$ & $5.83\times 10^{-4}$ \\
SDC-GON & $[6,4,6,8,1]$ & $\mathbf{2.95\times 10^{-4}}$ & $\mathbf{4.41\times 10^{-4}}$ \\
\hline
\end{tabular}
\end{table}

Table~\ref{tab:stokes} reports the quantitative results. The Stokes case shows that the proposed learning architecture can be applied to a vector-valued system. With the smaller architecture $[6,4,6,8,1]$, SDC-GON reaches a testing error of $4.41\times 10^{-4}$ in Table~\ref{tab:stokes}. This is lower than the $1.97\times 10^{-3}$ testing error of GON with the same smaller architecture, and it is also lower than the $5.83\times 10^{-4}$ testing error of the larger GON configuration $[6,12,24,12,1]$. These results indicate that the shared-backbone architecture and self-consistency regularization components of SDC-GON are independently effective for vector-valued systems, even when the analytical singular decomposition component is absent. The Stokes benchmark therefore provides evidence that the non-singular components of the proposed framework generalize beyond the scalar diffusion setting. A full Stokes-specific singular decomposition based on the Stokeslet tensor and its associated traction kernels, together with a complete stress-based boundary integral formulation, are left for future work.

\subsection{Ablation study}

An ablation study was conducted on the flat plate reaction diffusion benchmark to isolate the effects of the singular network, singular decomposition, learning rate decay, and consistency loss. Table~\ref{tab:ablation} reports eight configurations, including the baseline configuration without any of the proposed components. In this table, ``singular network'' refers to using the shared dual-head BsNN architecture without the analytical singular component, while ``singular decomposition'' refers to using the analytical $G_{\mathrm{sing}}$ term together with the shared architecture. Therefore, the singular network should be interpreted as an architectural change, whereas singular decomposition represents the analytical removal of the leading singular behavior.

\begin{table}[t]
\centering
\caption{Ablation study on the reaction diffusion benchmark.}
\label{tab:ablation}
\small
\begin{tabular}{llllllll}
\hline
Singular network & Singular decomposition & LR decay & Consistency loss & Training error & Testing error  \\
\hline
\xmark & \xmark & \xmark & \xmark & $3.80\times 10^{-4}$ & $2.52\times 10^{-3}$  \\
\xmark & \xmark & \xmark & \cmark & $4.60\times 10^{-4}$ & $7.40\times 10^{-4}$  \\
\xmark & \xmark & \cmark & \xmark & $4.80\times 10^{-4}$ & $6.30\times 10^{-4}$  \\
\xmark & \cmark & \xmark & \xmark & $3.70\times 10^{-4}$ & $6.90\times 10^{-4}$  \\
\cmark & \xmark & \xmark & \xmark & $3.70\times 10^{-3}$ & $1.10\times 10^{-3}$  \\
\cmark & \cmark & \xmark & \xmark & $4.80\times 10^{-4}$ & $5.60\times 10^{-4}$  \\
\cmark & \cmark & \cmark & \xmark & $3.70\times 10^{-4}$ & $4.20\times 10^{-4}$  \\
\cmark & \cmark & \cmark & \cmark & $\mathbf{1.00\times 10^{-4}}$ & $\mathbf{2.78\times 10^{-4}}$  \\
\hline
\end{tabular}
\end{table}

The ablation results show that each component contributes differently. The baseline configuration gives a testing error of $2.52\times 10^{-3}$. Adding consistency loss alone reduces the testing error to $7.40\times 10^{-4}$, learning rate decay alone gives $6.30\times 10^{-4}$, and singular decomposition alone gives $6.90\times 10^{-4}$. These results indicate that each component improves generalization relative to the baseline, but none of them alone reaches the performance of the full model.

The singular network alone gives a testing error of $1.10\times 10^{-3}$, which is lower than the baseline testing error of $2.52\times 10^{-3}$, but its training error increases from $3.80\times 10^{-4}$ to $3.70\times 10^{-3}$. This pattern suggests that the shared dual-head architecture changes the inductive bias of the model rather than simply improving data fitting. In particular, the shared backbone forces the scalar Green's function representation and the gradient representation to rely on common intermediate features. This coupling may reduce overfitting to the training samples and improve test performance, even though the model becomes less able to minimize the training loss when the analytical singular component is absent. Therefore, the singular network provides a useful architectural regularization effect, but it does not by itself solve the main approximation difficulty caused by the singular Green's function.

Combining the singular network with singular decomposition reduces the testing error further to $5.60\times 10^{-4}$. This confirms that the architectural change becomes more effective when the network is no longer required to learn the full singular kernel directly. Adding learning rate decay further reduces the testing error to $4.20\times 10^{-4}$, showing that optimization stability is important for the coupled scalar-gradient representation. The full SDC-GON configuration achieves the best result, with $1.00\times 10^{-4}$ training error and $2.78\times 10^{-4}$ testing error. These results show that the strongest performance is obtained when the architectural coupling, analytical singular decomposition, learning rate schedule, and smooth-gradient consistency loss are used together.

The ablation study was conducted on the flat plate reaction--diffusion benchmark because this case provides a representative diffusion-dominated problem while keeping the computational cost of repeated controlled experiments manageable. We did not perform a full component-wise ablation on every benchmark; therefore, the relative importance of singular decomposition, learning-rate decay, and consistency regularization should not be interpreted as identical across all geometries and operators. Instead, the flat plate ablation is intended to isolate the qualitative role of each component under one controlled setting.

The expected contribution of each component depends on the structure of the benchmark. The singular decomposition is most beneficial when the dominant singular behavior of the Green's function is known and can be removed analytically, as in the Poisson and diffusion--reaction cases. Learning-rate decay mainly improves optimization stability and late-stage refinement, and is therefore expected to have a more general effect across benchmarks. The consistency loss regularizes the relationship between the learned smooth correction and its gradient, which is especially relevant when boundary contributions depend on gradient information. For heterogeneous problems such as the pipe benchmark, the scalar constant-coefficient singular term is only an approximate singular decomposition, so its isolated contribution may differ from that observed in the flat plate case. Nevertheless, the full SDC-GON model improves over the corresponding GON baselines in the 3D heat conduction and heterogeneous pipe benchmarks, indicating that the combined design remains effective beyond the flat plate setting.

\section{Conclusions and Future Work}
\label{conclusion}

This work presented SDC-GON, a singular decomposition and consistency regularized Green's operator network for solving linear PDEs under varying boundary conditions and source terms. The method addresses two coupled limitations in Green's function learning: the singular behavior of the Green's function near the source point and the lack of explicit agreement between the learned Green's function and the gradient used in the boundary integral. By separating the analytical singular component from a smooth neural correction, SDC-GON changes the learning target from a singular kernel to a regular correction. By imposing self consistency between the smooth correction and its predicted gradient, the method improves the coherence of the quantities used in the integral solution representation. 

Across the two dimensional Poisson, three dimensional heat conduction, heterogeneous reaction diffusion, and Stokes benchmarks, SDC-GON consistently achieved lower testing errors than the compared baselines. On the heterogeneous pipe benchmark, SDC-GON obtained a testing error of $3.70\times10^{-4}$ with a smaller network architecture, compared with $9.60\times10^{-4}$ for the same-width GON baseline and $4.63\times10^{-4}$ for a larger GON configuration. This result shows that the proposed structural changes can be more effective than simply increasing model size. The Stokes case demonstrates that the shared-backbone architecture and self-consistency regularization components of SDC-GON are independently effective for vector-valued systems, even when the analytical singular decomposition component is not applied. With the smaller architecture $[6,4,6,8,1]$, SDC-GON achieved a testing error of $4.41\times10^{-4}$, lower than both the same-width GON baseline ($1.97\times10^{-3}$) and the larger GON configuration ($5.83\times10^{-4}$). The Stokes gains therefore provide evidence that the non-singular components of the framework generalize beyond the scalar diffusion benchmarks. The ablation study confirms that singular decomposition and the warmup cosine schedule are the main contributors to accuracy and stability, while consistency regularization provides additional improvement.

The main advantage of SDC-GON is that it preserves the reusable response representation of Green's function based operator learning while making the learned representation more stable and physically interpretable. The singular decomposition embeds the leading local behavior of the Green's function analytically and assigns the neural network only the smoother remainder. The self consistency loss is also distinct from a standard PDE residual loss: it is imposed directly on the internal Green's function representation by enforcing agreement between the smooth correction and the gradient required by the boundary integral. This distinction is important because an unconstrained gradient head may still produce a solution with low reconstruction error while remaining inconsistent with the scalar Green's function approximation. The warmup cosine learning rate schedule complements these components by stabilizing the coupled optimization of the solution and consistency losses.

The current framework also has several limitations that suggest directions for future work. First, the present study focuses on linear diffusion-dominated operators in bounded domains. Extension to nonlinear PDEs would require either iterative linearization or a modified learning architecture. For example, SDC-GON could be combined with neural-network-assisted Picard or Newton iterations for nonlinear boundary value problems, where each iteration solves a linearized problem using a learned Green's function operator, in a manner related to DeepGreen~\cite{gin2021deepgreen}. Second, the singular component used in this work is based on the free-space diffusion kernel. Although this choice is effective for the Poisson and diffusion--reaction problems studied here, it may not directly apply to convection-dominated, strongly anisotropic, nonlocal, or tightly coupled multiphysics operators, where the local fundamental solution differs from the isotropic diffusion kernel. Future work should therefore develop operator-specific and coefficient-adapted singular decompositions for these broader PDE classes. For the Stokes system in particular, this would require a Stokes-specific decomposition based on the Stokeslet tensor and its associated traction kernels, together with a complete stress-based boundary integral formulation. Third, the current consistency loss enforces pointwise agreement between the learned smooth correction and its gradient. Future work could further develop integral-level consistency constraints that connect the pointwise correction and gradient consistency more directly to the resulting volume and boundary integral contributions. These directions would help evaluate whether combining analytical singularity treatment with learned smooth corrections can serve as a general design principle for physics-consistent operator learning across broader classes of PDEs.

\section*{Acknowledgements}

This research was supported by funding from Saskatchewan Polytechnic, Faculty of Digital Innovation, Arts and Sciences and the 2026 Applied Research Release Time (ARRT) Fund from Saskatchewan Polytechnic.

\section*{Declaration of Competing Interest}

The authors declare that they have no known competing financial interests or personal relationships that could have appeared to influence the work reported in this paper.

\section*{Data availability}
The data will be shared upon reasonable request.

\section*{Code Availability}

A GitHub repository will be shared upon request.

\section*{Author Contributions}

Yingchao Huang: Conceptualization, methodology, data curation, formal analysis, visualization, writing the original draft. 
Xin Wang: Conceptualization, validation, visualization, review and editing. 
Shanshan Yao: Validation, formal analysis, and visualization. 
Fanhua Zeng: Validation, review and editing. 
Wei Peng: Visualization, review and editing.

\bibliography{sdc_gon}

\end{document}